\documentclass{article}
\usepackage[utf8]{inputenc}
\usepackage{graphicx}

\title{Multi-node environment strategy for Parallel Deterministic Multi-Objective Fractal Decomposition }
\author{L\'{e}o Souquet, Amir Nakib \\
Universit\'{e} Paris-Est, Laboratoire LISSI, \\ 122 Rue Paul Armangot, 94400\\ Vitry sur Seine, France\\
}
\date{August 2019}
	
\begin{document}
\maketitle

	\begin{abstract}
		This paper presents a new implementation of deterministic multiobjective (MO) optimization called "Multiobjective Fractal Decomposition Algorithm" (Mo-FDA). The original algorithm was designed for mono-objective large-scale continuous optimization problems. It is based on a "divide-and-conquer" strategy and a geometric fractal decomposition of the search space using hyperspheres. Then, to deal with MO problems a scalarization approach is used. In this work, a new approach has been developed on a multi-node environment using containers. The performance of Mo-FDA was compared to state of the art algorithms from the literature on classical benchmark of multi-objective optimization.
	\end{abstract}
	
	\section{Introduction}
	
	In multiobjective optimization problems (MOP) the goal is to optimize at least two objective functions. This paper deals with these problems by using a new decomposition-based algorithm called: "Fractal geometric decomposition base algorithm" (FDA). It is a deterministic metaheuristic developed to solve large-scale continuous optimization problems \cite{Nakib2017}. It can be noticed, that we call large scale problems those having the dimension greater than 1000.
	In this research, we are interested in using FDA to deal with MOPs because in the literature decomposition based algorithms have been with more less success applied to solve these problems, their main problem is related to their complexity. In this work, the goal is to deal with this complexity problem by keeping the same level of efficiency.
	FDA is based on "divide-and-conquer" paradigm where the sub-regions are hyperspheres rather than hypercubes on classical approaches. 
	In order to identify the Pareto optimal solutions, we propose to extend FDA using the scalarization approach. We called the proposed algorithm Mo-FDA. 
	This new approach has been developed to benefit from a multi-node environment to improve the computational time taken to solved MOPs problems. This chosen architecture takes profit from containers, light weight virtual machines that are design to run a specific task only. A single machine can host many more containers than regular virtual instances.

	The rest of the of paper is organized as follow. The next Section \ref{mofda} presents a description of the proposed algorithm. Section \ref{architecture} presents the chosen architecture. In Section \ref{result} the obtained results and a comparison to the competing methods are presented. Finally, Section \ref{conclusion} presents the future work.

	\section{Proposed Mo-FDA}
	\label{mofda}

	Fractal Decomposition Algorithm uses a "Divide-and-Conquer" strategy across the search space to find the global optimum, when it exists, or the best local optimum. The hypercubes are the most used forms in the literature. However, this geometrical form is not adapted to solve large scale problems or high dimensional problems because the number of vertices increases exponentially. FDA~\cite{Nakib2017} uses hyperspheres to divide the search domain as this geometrical object scales well as the dimension increases allowing FDA to solve large scale problems. In addition, the fractal aspect of FDA is a reference to the fact that the search domain is decomposed using the same pattern at each level until the maximum fractal depth \textit{k} (fixed by the user). While searching for the optimum, FDA uses 3 phases: initialization phase; 1) exploration phase, 2) and exploitation phase 3). During the initialization phase, at level 0, the current hypersphere is decomposed into $2 \times D$ sub-hyperspheres with $D$ being the problem's dimension.
	
	Once the initialization phase is completed, FDA starts the exploration phase  to identify the sub-hypersphere that potentially contains the global optimum or best local optimum (or global optimum if it is known), to decompose it again in $2 \times D$ sub-hyperspheres. This operation is repeated until the maximum fractal depth, $k$ is reached. $k$ has been experimentally determined and set to 5. When the $k-level$ is reached, FDA enters in the exploitation phase. 
	The aim of this phase is to find the best local optimum inside the current sub-hyperspheres. This procedure is called \textit{Intensification local search (ILS)}. Each instance of ILS starts at the center of the sub-hypersphere being exploited. This local search is moving along each dimension, evaluating three points on each one and only the best is considered for the following dimension (more details are in \cite{Nakib2017}). Then, the second \textit{k-level sub-hypersphere} is exploited using ILS. This process will stop when the maximum number of evaluations is reached. If all \textit{k-level sub-hypersphere} have been exploited without FDA stopping, it backtracks to decompose the second best hypersphere at the level $k-1$.
	
	In a multi-objective problem, different objective functions are being optimized at the same time. Scalarization techniques allows to combine the different objective functions into one, allowing the approach to solve it as a mono-objective problem. Different scalarization methods were proposed such as Weighted Sum and Weighted Tchebycheff Method \cite{Miettinen2008}. In this work, Tchebycheff approach were considered for Mo-FDA and is defined in equation \ref{tchetche}:
	\begin{equation}
	\label{tchetche}
	\begin{array}{rrclcl}
	Minimize & 
	\displaystyle  \max_{i= 1,...,k} &
	
	\multicolumn{3}{l}{[\omega_{i} (f_i(x) - z_i^*)]}\\
	\textrm{Subject to} & x \in \textit{X}\\
	\end{array}
	\end{equation}
	
	with $k$ the number of objective functions to optimise and $z^*_i$ the optimum of function $f_i$. In addition, the sum of weights $\omega_{i}$ must be equal to 1. In Mo-FDA, $n$ different mono-objective problems are solved with different combination of weights $\omega_i$. As each combination produces one solution, Mo-FDA produces a Pareto-Front (PF) composed of $n$ points. To improve the
	speed at which Mo-FDA solves a multi-objective problem, the $n$ independent instances are launched simultaneously using containers. The overhead produced by the different containers' management can be neglected compared to the benefit from the parallel implementation of the $n$ instances. Furthermore, one can see that the algorithm is running on a multi-node environment without having to change the implementation.
	
	
	
	\section{Proposed strategy}
	\label{architecture}
	
	 Using this technique, to obtain $n$ points in the Pareto-Front(PF), the algorithm will be launched $n$ times with $n$ variation of the weights $\omega$ as showed in equation \ref{tchetche} leading a significant increase in computational time. Then, the idea to overcome this problem is to design a multi-node architecture. 
	 
	 The goal is to have each node finding one point corresponding to one combination of the weights $\omega$ and combine all their results to build the full PF. The challenge behind this architecture is that the computing resources needed increase with the size of the Pareto-Front. For instance, if $n = 100$ points, it means that 100 nodes would be required, hence 100 different computers (or virtual machines), which can be seen as an oversized architecture. To tackle this important issue, we proposed a strategy based on using $containers$. They are significantly lighter than virtual machines as they all share the same operating system kernel. This way, a single machine can host more containers than virtual machines. This architecture significantly less complex and allows to benefit from multi-node approaches while developing it on a limited number of hosts. 
	
	In parallel to this multiple containers running on a single machine approach, a multi-threaded architecture was also studied. However, the threads are part of the same main process was not compatible with the desired architecture to have $n$ independent instances of the algorithm.	In addition containers can be deployed on multiple different physical machines seamlessly, without having to change the structure of our algorithm. This cannot be achieved if Mo-FDA was developed using multi-threads.

	\section{Results and Discussion}
	\label{result}
	
	In order to test the performance of Mo-FDA a set of 8 functions, 5 from the ZDT family problems \cite{Deb2005} and 3 from the DTLZ sets \cite{DTLZ} have been used. The results obtained were compared to the well known algorithms NSGA-II, NSGA-III and MEOA/D as well as state-of-the-art approaches GWASFGA \cite{GWASFGA}, and CDG \cite{CDG}. To conduct the different experiments, jMetal 5.0 \cite{jMetal}, a popular Java-based framework in the literature has been used. The principal experiments settings described in \cite{JIANG2014125}.

	
	In the context of multi-objective problems, many metrics can be used as discussed in \cite{Metrics}. In order to have a better overview of the performances of Mo-FDA compared to other algorithms, we have selected four different metrics. The first one, the Hypervolume metric. It measures the size of the portion of the objective space that is dominated by an approximation set. The Generational Distance metric (GD) computes the average distance from a set of solutions obtained by an algorithm to the true Pareto-Front. The Inverted generational distance (IGD), measures both convergence and diversity by computing the distance from each point known in the true Pareto-Front to each point of a set of solutions found by the executed algorithm. The Spread metric measures the extent of the spread achieved among the obtained solutions. It is important to notice that the goal is to maximize the first metric and to minimize the others.

\begin{figure}%
		\centering
		{\includegraphics[width=5cm]{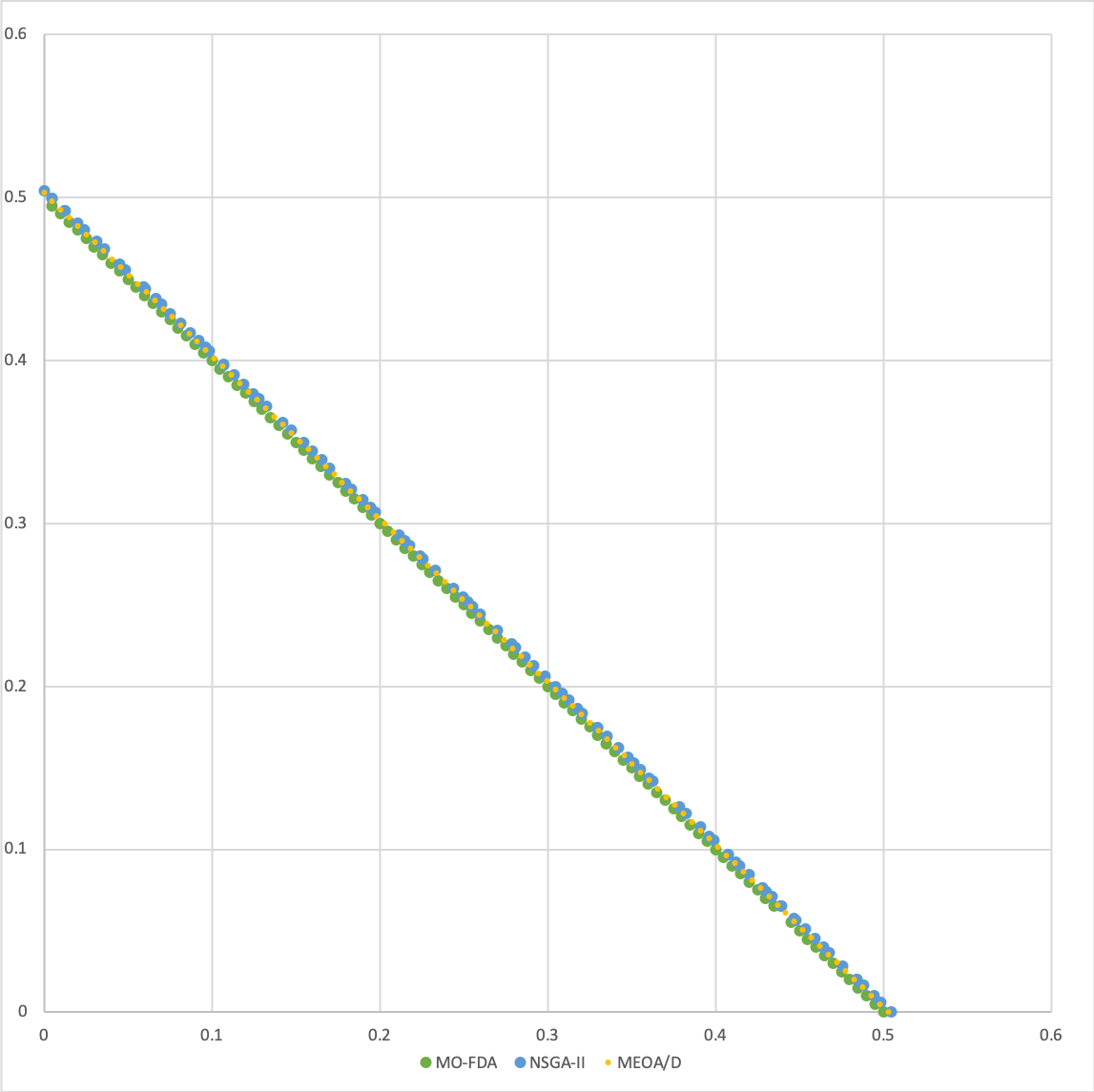} }%
		\qquad
		{\includegraphics[width=5cm]{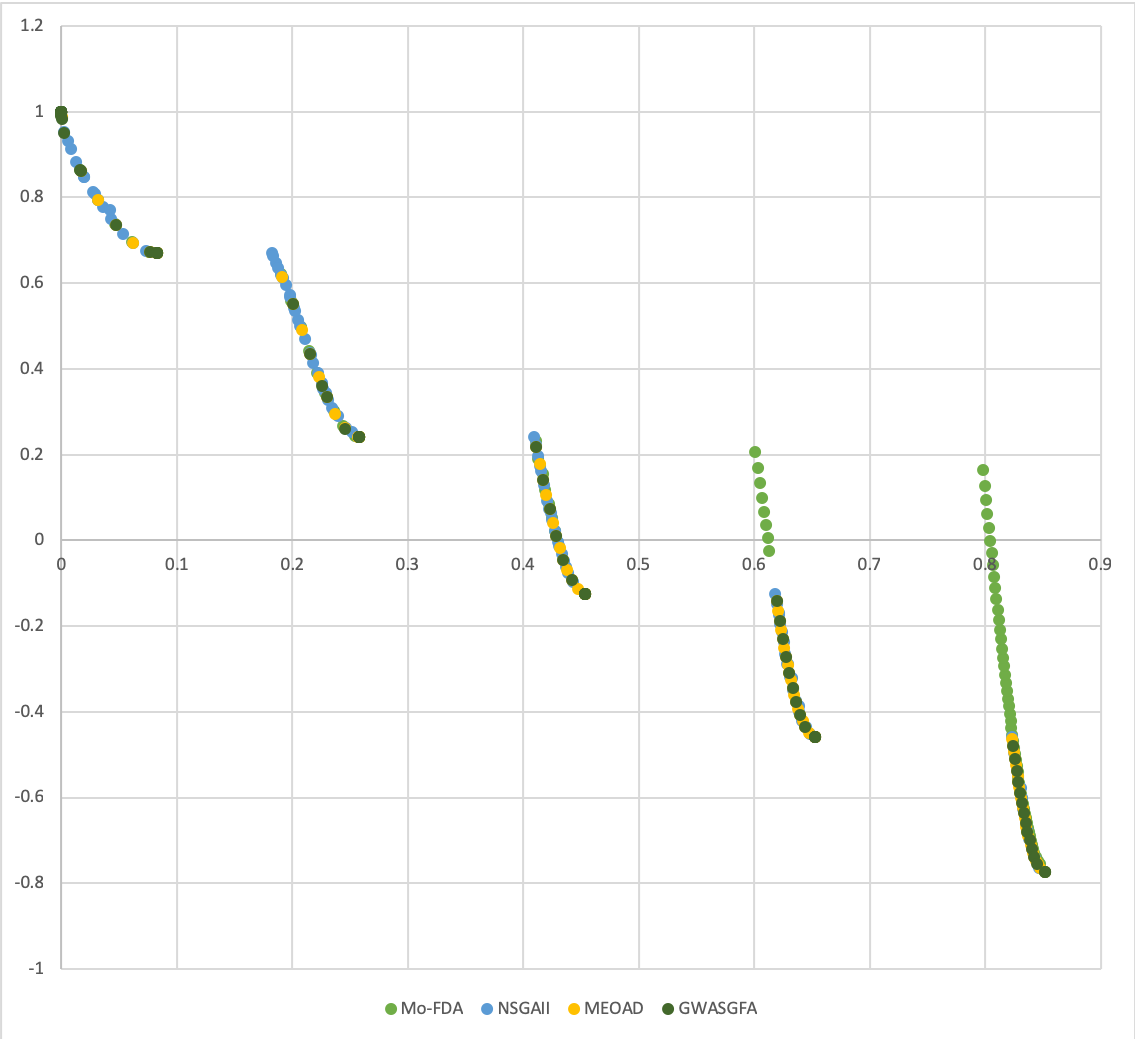} }%
		\caption{Pareto-Front of DTLZ1 (on the left) and of ZDT3 (on the right)}%
		\label{ParetoFront}%
	\end{figure}

	Moreover, to compare the results obtained by the different algorithms we used the Friedman Rank sum method and the obtained results are presented in Table \ref{Results}. One can see that Mo-FDA is the most efficient algorithm among three metrics out of four. Looking at the other algorithms, MEOAD/D is efficient on the IGD and Spread but not on the GD and the Hypervolume. Consequently, the importance of using multiple criteria highlights strengths and weakness of each algorithm.  Figure \ref{ParetoFront} shows the Pareto-Front found by Mo-FDA and the best algorithms on two functions, DTLZ1 and ZDT3. Functions where Mo-FDA performs the best and the worst respectively. 
	
In addition to the precision of the algorithm we have also measured the new time performance of the proposed architecture. Over the 8 different functions tested, on average, the computation time required to solve one function at dimension with our architecture \textit{D=30} is 0.8 seconds on a single host. Using two separate hosts, the time was reduced to 0.5 seconds. This shows that our architecture is scalable and flexible without having to change the structure of the algorithm nor the implementation itself. All experimentations have been conducted using the following characteristics: Mo-FDA has been developed in Python and the nodes have a processor Intel Xeon Platinum 8000 with 144GB of RAM.

	\begin{table}
		\begin{center}
			\begin{tabular}{l r r r r r r}
				\textbf{Algorithms} & \textbf{Mo-FDA} & \textbf{NSGA-II} & \textbf{NSGA-III} & \textbf{MEOA/D} & \textbf{GWASGFA} & \textbf{CDG} \\
				\hline
				Hypervolume & 1.875 (1) & 2.25 (2) & 5.625 (6)& 2.625 (3)& 4.125 (4)& 4.25 (5)\\
				GD & 2.25 (1) & 2.875 (2) & 4.875 (6) & 3.25 (4) & 3.125 (3)	 & 4.625 (5) \\
				IGD & 2.125 (2) & 2.25 (3) & 5.5 (6) & 2 (1) & 4.25 (4) & 4.875 (5) \\
				Spread & 1.75 (1) & 3.5 (3) & 4.25 (4) & 1.75 (1) & 4.25 (4) & 5.5 (6)\\
				\hline
			\end{tabular}
		\end{center}
		\caption{Ranking using Friedman Rank sum (and their global rank) of all algorithms on the different metrics for all tested functions.}
		\label{Results}
	\end{table}
	
	\section{Conclusion and future work}
	\label{conclusion}
	
		In conclusion, Mo-FDA was tested on 8 different functions and compared to 5 other well regarded and state-of-the-art metaheuristics. Its performances to find good Pareto-front is proved using four popular metrics in the literature. In addition, the multi-node architecture using containers improves the flexibility and scalability of the approach while reducing the computing resources needed. For future work, we aim to adapt Mo-FDA to many-objectives problems and apply it to a real world problems.

	\bibliography{example}
	\bibliographystyle{plain}

\end{document}